\documentclass[%
reprint,
superscriptaddress,
amsmath,amssymb,
aps,
prb
floatfix,
showpacs,
]{revtex4-2}

\usepackage{graphicx}
\usepackage{dcolumn}
\usepackage{bm}
\usepackage{hyperref}
\usepackage{ulem}

\usepackage[english]{babel}
\usepackage[utf8]{inputenc}
\usepackage{amssymb,amsfonts,amsmath,mathtools,mathrsfs}
\usepackage{relsize}
\usepackage{mdframed}
\usepackage[margin=0.6in]{geometry}
\usepackage{enumitem}
\usepackage{graphicx}
\usepackage{float}
\usepackage{url,hyperref}
\usepackage[table,usenames,dvipsnames]{xcolor}
\usepackage{gensymb}
\usepackage{grffile}

\hypersetup{
	colorlinks=true,
	linkcolor=Blue,
	filecolor=magenta,      
	urlcolor=Blue,
	citecolor=Blue,
}
\usepackage[bottom]{footmisc}
\usepackage[capitalize]{cleveref}
\usepackage{textcomp}
\usepackage{braket} 

\setcitestyle{line}

\usepackage{amsmath,amssymb,wasysym}

\makeatletter
\def\@fnsymbol#1{\ensuremath{\ifcase#1\or \dagger\or *\or \ddagger\or
   \mathsection\or \mathparagraph\or \|\or **\or \dagger\dagger
   \or \ddagger\ddagger \else\@ctrerr\fi}}
    \makeatother

\begin{document}


\title{Weak anti-localization and spin-momentum locking in topological insulator Ta$_2$Ni$_3$Te$_5$}

\author{Prabuddha Kant Mishra}
\address{Department of Chemistry, Indian Institute of Technology Delhi, New Delhi 110016, India}
\author{Soumen Ash} 
\altaffiliation{Current Affiliation: Leibniz Institute for Solid State and Materials Research Dresden, 01069 Dresden, Germany.}
\address{Department of Chemistry, Indian Institute of Technology Delhi, New Delhi 110016, India}
\author{Ashok Kumar Ganguli}\email[E-mail: ]{ashok@chemistry.iitd.ac.in}
\address{Department of Chemistry, Indian Institute of Technology Delhi, New Delhi 110016, India}
\address{Department of Materials Science and Engineering, Indian Institute of Technology Delhi, New Delhi 110016, India}

\begin{abstract}
 
We report the synthesis, structural characterization, and investigation of electrical transport, magnetic and specific heat properties of bulk semiconducting layered material Ta$_2$Ni$_3$Te$_5$. Ta$_2$Ni$_3$Te$_5$ crystallizes in the centrosymmetric orthorhombic structure with space group $Pnma$. Temperature-dependent resistivity shows a transition from semiconducting to metallic nature below $\approx$ 7 K. Low-temperature magnetotransport studies show large magnetoresistance with the signature of weak anti-localization (WAL) effect. The magnetoconductivity data has been used to explore the origin of the WAL effect, extract relevant parameters, and study their variation with temperature. The presence of significant electron-phonon interaction is evident from the $MR$ $vs.$ $B/R$ plot (Kohler's plot). Isothermal field-dependent magnetization studies show Berry paramagnetism as the signature for spin-orbit coupling-induced spin-momentum locking.
%
Observation of WAL effect and spin-momentum locking phenomenon demonstrate Ta$_2$Ni$_3$Te$_5$ as a promising low dimensional material for quantum spin Hall insulator-based applications.

\end{abstract}



\maketitle

\section{Introduction}

The physics of 2D materials has been grasping intense interest due to their non-trivial physical properties and applications. Topological perspectives of metal polychalcogenides have been studied for a detailed understanding of electronic band structure and their reflections in magnetotransport and magnetic properties\cite{Zhang2012, Zhang2013, Sahu2018, Zhang2011, Taskin2009, Hooda2018}. Taking spin-orbit coupling into consideration, the origin of the nontrivial topology of electronic band structure has been predicted for materials \cite{Olsen2019}. Quantum spin Hall insulators (QSHIs) are promising nontrivial topological candidates, having an inverted band structure close to the Fermi surface, thus consisting insulating 2D layer and conducting edge states \cite{Yong2013, Guo2021, Wu2016, Hasan2010}. Due to Kramer's degeneracy, these counterpropagating helical edge states feature pairs of conducting states with the opposite spin-polarization, which are protected by time-reversal symmetry and get gapped out after applying external magnetic field \cite{Markus2007, Qi2011, Hasan2010}. Theoretically, for centrosymmetric materials, the electronic band structure follows the relation under space inversion, $E (K, \uparrow) = E (-K, \uparrow)$, while it is violated at the surface of the material, and bands get split in the momentum space \cite{Dario2015} due to Rashba SOI, thus facilitating the spin-momentum locking \cite{Koga2002, Takasuna2017}. Spin-dependence of the electronic properties of materials introduced spintronics, an exotic field of research with vast applications in sensors and quantum computing \cite{Nayak2008a}. 

Exploring novel materials with strong spin-orbit coupling (SOC), that can provide unconventional and rich physics is an exciting endeavor. Such materials have been extensively studied through magnetotransport measurements in thin films \cite{Zhang2013}, single crystals \cite{Pavlosiuk2016, Lu2017, Hou2015, Taskin2009, Chiu2013}, and polycrystalline phases \cite{Zhang2012, Sahu2018, Zhang2011, Zhang2016, Shekhar2012}. The elements with a high atomic number are expected to have a substantial amount of SOC (SOC $\propto   Z^4$) \cite{Shanavas2014}, which plays a key role in the realization of QSHIs \cite{Gang2018}. Ta$_2$Ni$_3$Te$_5$, a metal-rich telluride ($Z_{Ta}=73$, $Z_{Ni}=28$ $Z_{Te}=52$) is therefore expected to have significant SOC and is an interesting system to investigate with magnetotransport studies. Synthesis and structural characterization of Ta$_2$Ni$_3$Te$_5$ was first discussed by W. Tremel \cite{Tremel1991}. Ta$_2$Ni$_3$Te$_5$ is a layered material, having weak van der Waals interactions between the layers. It crystallizes in a centrosymmetric orthorhombic crystal structure with space group $Pnma$. The electronic band structure of QSHI candidate Ta$_2$Ni$_3$Te$_5$ and its Pd analog has recently been studied by Guo \textit{et al.} \cite{Guo2021}. Besides, its structural similarity to Ta$_2$Pd$_3$Te$_5$, Ta$_2$Ni$_3$Te$_5$ need to be explored to correlate the effect of SOC in these ternary metal chalcogenides.
Nontrivial Chern mirror number indicates the presence of inverted band structure in Ta$_2$Ni$_3$Te$_5$ 3D crystal. From the theoretical study, the mono-layer of Ta$_2$Ni$_3$Te$_5$ is found to be a narrow-gapped semiconductor with the existence of non-trivial topological edge states \cite{Guo2021, Guo2022}. The pressure-induced metallization and appearance of superconductivity ($T_c$ $\sim$ 0.4 K, $P_c$ = 21.3 GPa) in Ta$_2$Ni$_3$Te$_5$ has been recently studied by Yang \textit{et al.} \cite{Yang2023}. In the case of isostructural Ta$_2$Pd$_3$Te$_5$, the edge states have been probed with scanning tunneling microscopy (STM), and angle-resolved photoemission spectroscopy (ARPES) measurements \cite{Wang2021}. Recently, Tian \textit{et al.} have studied the magnetotransport properties of Ta$_2$Pd$_3$Te$_5$ and observed weak anti-localization (WAL) to weak localization (WL) crossover as a function of temperature \cite{Tian2022}. 

Despite the theoretical understanding of the nontrivial band topology of Ta$_2$Ni$_3$Te$_5$, the experimental advancement in this direction is still lacking, and therefore, the physical properties of Ta$_2$Ni$_3$Te$_5$ are to be explored in detail. In this study, we report successful synthesis of highly pure polycrystalline samples of Ta$_2$Ni$_3$Te$_5$ by solid-state reaction technique. Detailed measurements have been carried out to study the magnetotransport, magnetic and thermal properties of Ta$_2$Ni$_3$Te$_5$. Ta$_2$Ni$_3$Te$_5$ is found to be a narrow band gap semiconductor at room temperature, while its low-temperature properties are driven by the interactions due to SOC. Based on the phase coherence effects in magnetoconductance, we can infer the existence of significant electron-phonon interaction in this bulk material. The magnetotransport study shows a weak anti-localization effect at low temperatures and low applied magnetic field regime. The isothermal field-dependent magnetization data show the existence of spin-momentum-locked electronic states. WAL effect and spin-momentum locking are signatures for the nontrivial topology of the electronic band structure, making Ta$_2$Ni$_3$Te$_5$ a promising candidate to investigate exotic quantum phenomena for QSHI-based applications.

\section{Experimental Details}

Polycrystalline samples with a nominal composition of Ta$_2$Ni$_3$Te$_5$ were synthesized by reacting Ta, Ni, and Te in elemental form. The stoichiometric amounts of reactants were mixed well, sealed in the evacuated quartz tube, and heated at 1173 K for 7 days. The heat-treated samples were then reground, pelletized, and sintered at 1173 K for another 7 days to achieve better phase homogeneity. The phase purity of the sample was verified by powder x-ray diffraction (XRD) technique using Bruker D8 Advance diffractometer with Cu-$K\alpha$ radiation. The structural refinement of the room-temperature powder x-ray diffraction data was carried out using the Rietveld method with TOPAS software package \cite{Topas}. The compositional analysis and elemental mapping of the sample were carried out in the field emission scanning electron microscope (FESEM, JEOL JSM-7800F). The electrical resistivity and magnetotransport measurements were performed using four probe method with AC phase-sensitive lock-in technique. The data were collected for the temperature range of 2-300 K and magnetic fields up to 14 T using a physical property measurement system (PPMS, Cryogenic Limited, UK). Field-dependent and temperature-dependent magnetic measurements and specific heat measurement were carried out in the same PPMS. The details of various parameters of measurement conditions for transport measurements and magnetic measurement protocols are mentioned in the supplementary.

\section{structural studies}
Fig. \hyperref[Fig. 1]{1}(a) shows the Rietveld refinement of room temperature powder x-ray diffraction pattern of Ta$_2$Ni$_3$Te$_5$. It crystallizes in the centrosymmetric orthorhombic structure with space group $Pnma$, and the refined lattice parameters were obtained as $a$ = 13.881(1) \r{A}, $b$ = 3.661(1) \r{A} and $c$ = 17.782(1) \r{A}, consistent with the previous report \cite{Tremel1991}. The refined structural parameters are given in Table 1. From the powder XRD pattern, it can be observed that there is a preferred orientation along the $(h00)$ crystallographic direction, which indicates the layered nature of the crystallites. The layered nature of the material is also evident from the images recorded in the scanning electron microscope (SEM), shown in Fig. \hyperref[Fig. 1]{1}(b). Elemental mapping shows uniform distribution of all the elements. Energy dispersive X-ray (EDX) composition analysis provides the elemental ratio (Ta:Ni:Te = 2:2.7:5) close to the nominal composition.
Fig. \hyperref[Fig. 1]{1}(c) depicts the crystal structure of  Ta$_2$Ni$_3$Te$_5$. The crystal structure can be viewed as the clusters of Ta$_2$Ni$_2$ in which each Ni atom is tetrahedrally coordinated by \textit{four} Te atoms. The layered structure can be seen as made of quasi-one-dimensional chains along the b-axis consisting of edge-shared square pyramidal polyhedra of TaTe$_5$. Packing of TaTe$_5$ generates two types of voids in the lattice system, and filling of these voids with Ni leads to Ta$_2$Ni$_3$Te$_5$. These Ta$_2$Ni$_2$, NiTe$_4$, and TaTe$_5$ polyhedron units are depicted in Fig. \hyperref[Fig. 1]{1}(c) in their respective positions in the unit cell. 

\begin{figure}
\includegraphics[width= 1.0\columnwidth,angle=0,clip=true]{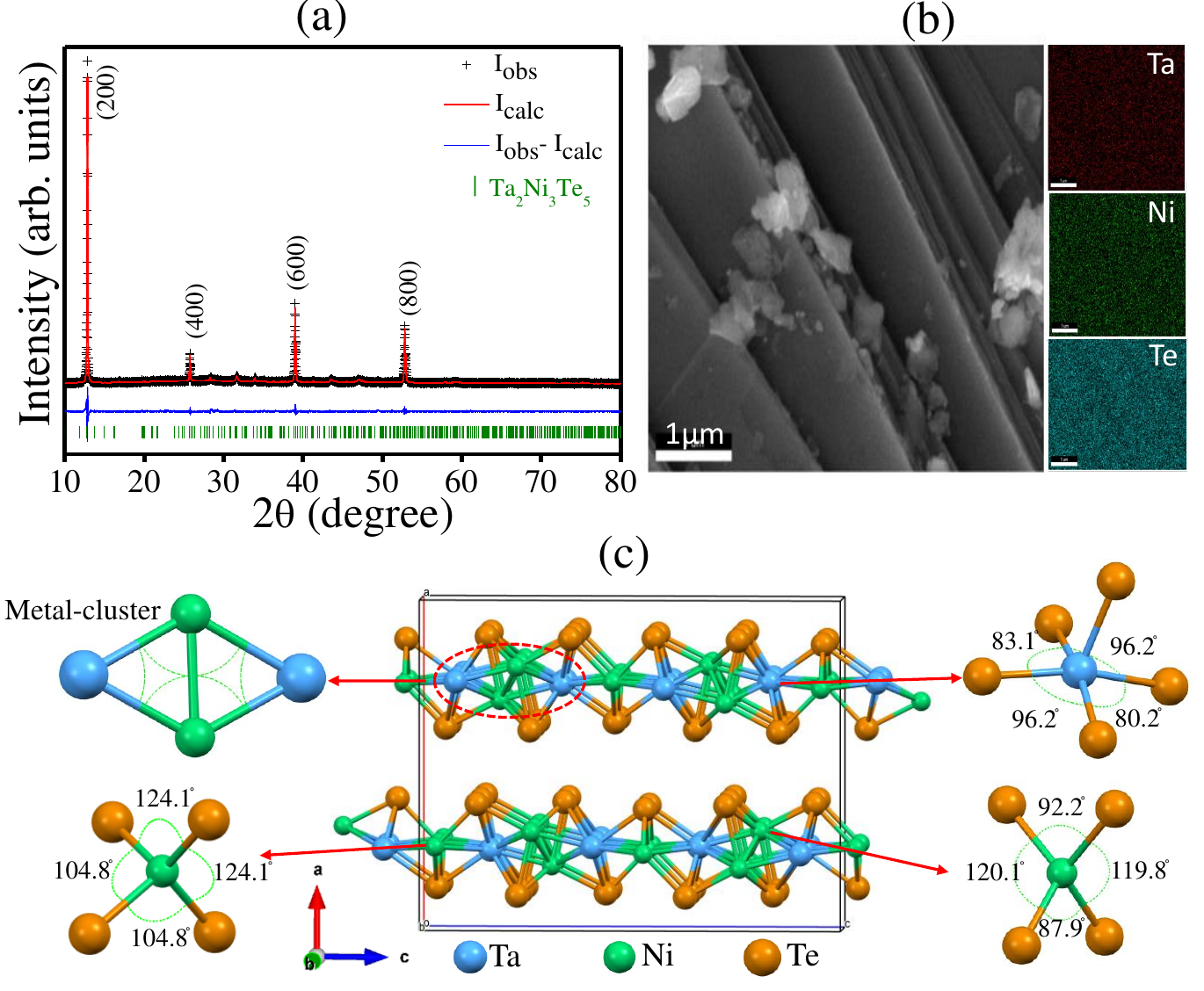}
\caption{(Color online) (a) Rietveld refinement of room temperature powder XRD data of polycrystalline bulk Ta$_2$Ni$_3$Te$_5$. Vertical bars represent the 2$\theta$ positions for allowed Bragg's reflections of Ta$_2$Ni$_3$Te$_5$. The blue line is the difference between the observed and the fitted patterns. (b) Stacked layers of material, visible in SEM image of the sample, along with elemental mapping (c) Crystal structure of Ta$_2$Ni$_3$Te$_5$, layers stacked along a-axis. Various metal-centered polyhedron and metal-cluster of Ta$_2$Ni$_2$ are depicted on both sides of the unit cell.}
\label{Fig. 1}
\end{figure}


\begin{table}[h]
\scriptsize\addtolength{\tabcolsep}{-1pt}
\scriptsize
\caption{\label{table1}Refined structural parameters of Ta$_2$Ni$_3$Te$_5$ in their respective units.}
\centering
\begin{ruledtabular}
\begin{tabular}{l c c c c c r}
Ta$_2$Ni$_3$Te$_5$ \\
Space group: & $Pnma$ \\
{\it a }  &13.881(1) \r{A}\\  
{\it b }  &3.661(1) \r{A}\\
{\it c } & 17.782(1) \r{A}\\
$\chi^2$ & 1.25 $\%$\\
\hline

Atom & Site & x & y & z & Occu. & B$_{eq}$ (\r{A}$^2$) \\
Ta(1) & 4c & 0.75(3) & 0.75 & 0.31(4) & 1 & 2.4(2) \\
Ta(2) & 4c & 0.71(4) & 0.25 & 0.07(2) & 1 & 2.6(3) \\
Ni(1) & 4c & 0.78(2) & 0.25 & 0.44(4) & 1 & 3.4(1) \\
Ni(2) & 4c & 0.80(2) & 0.25 & 0.24(4) & 1 & 3.8(3) \\
Ni(3) & 4c & 0.66(3) & 0.75 & 0.16(2) & 1 & 4.8(1) \\
Te(1) & 4c & 0.58(3) & 0.25 & 0.35(1) & 1 & 4.2(2) \\
Te(2) & 4c & 0.87(4) & 0.25 & 0.63(2) & 1 & 3.2(2) \\
Te(3) & 4c & 0.59(2) & -0.25 & 0.35(2) & 1 & 3.3(1) \\
Te(4) & 4c & 0.88(2) & -0.25 & 0.04(2) & 1 & 3.3(4) \\
Te(5) & 4c & 0.96(1) & 0.25 & 0.48(2) & 1 & 3.8(3) \\
\end{tabular}
\end{ruledtabular}
\end{table}


\section{Transport Properties}

\begin{figure}[t!]
\includegraphics[width=1.0 \columnwidth,angle=0,clip=true]{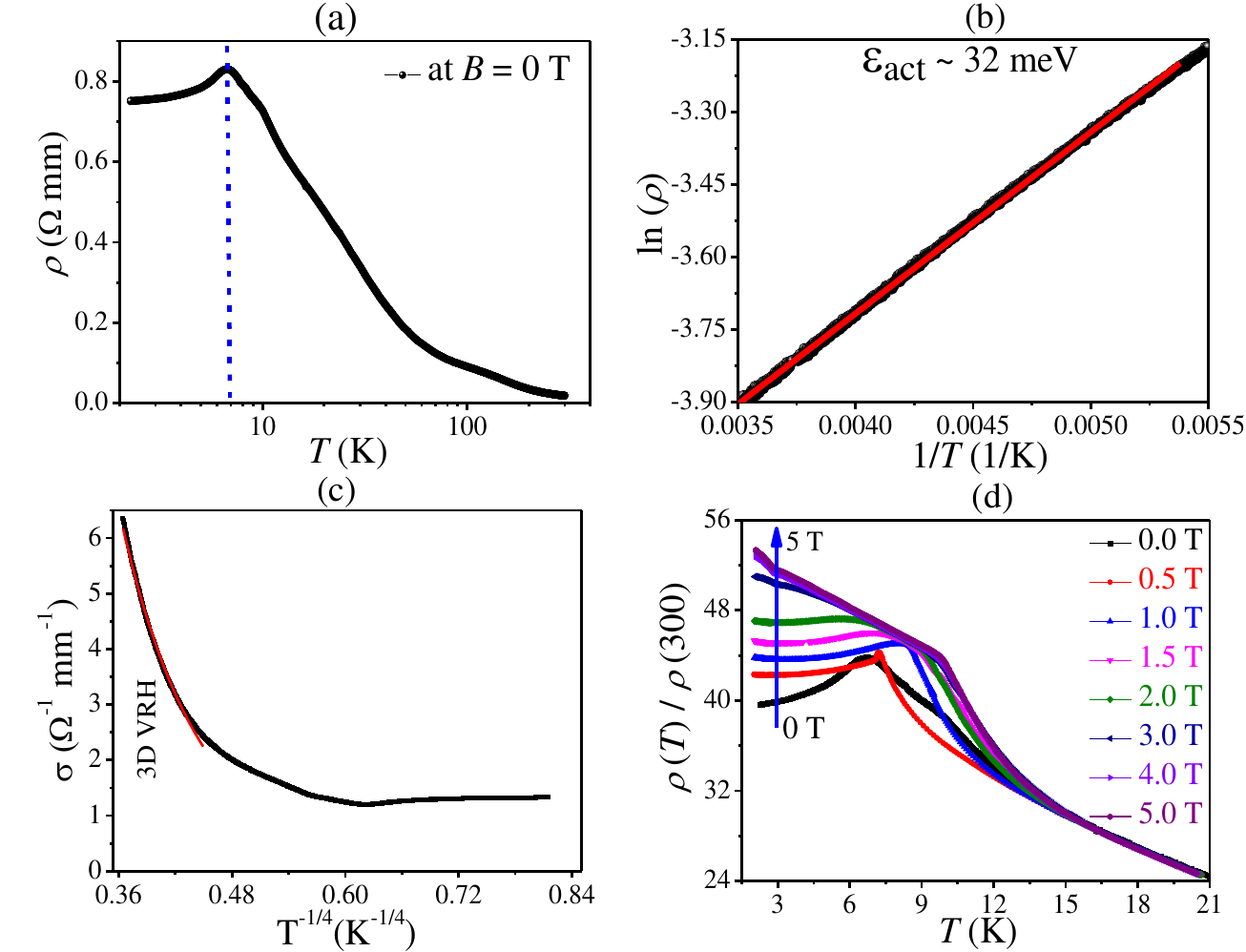}
\caption{(Color online) (a) Resistivity as a function of temperature (logarithmic scale) showing semiconducting to semimetallic transition for polycrystalline Ta$_2$Ni$_3$Te$_5$ at around 7 K, indicated by a dotted line. (b) Arrhenius fit for high-temperature data ( 180-300 K), shown by solid red line. (c) Electrical conductivity as a function of $T^{-1/4}$. The VRH fit for conductivity is shown in the solid red line, having deviation from data at low temperature. (d) The resistivity, normalized w.r.t. resistivity at 300 K ($\rho(300)$), as a function of temperature at various applied magnetic field perpendicular to the current direction.}
\label{Fig. 2}
\end{figure}

Fig. \hyperref[Fig. 2]{2}(a) shows the temperature dependence of the resistivity $\rho(T)$ data studied under zero magnetic field. As discernible from the figure, when the temperature is decreased from 300 K to 100 K, the resistivity increases monotonically, indicating semiconducting behavior. For the clear understanding of the temperature dependence of resistivity it have been discussed in the three separate regimes as (i) high-temperature regime (250 K $<T<$ 300 K) and (ii) mid-temperature regime (25 K $<T<$ 60 K) (iii) low-temperature regime (2 K $<T<$ 25 K). In the high-temperature region, temperature-dependent resistivity can be fit with the Arrhenius model $\rho(T)$ $\propto$ $exp(\varepsilon_{act}/k_BT)$, where $k_B$ and $\varepsilon_{act}$ are the Boltzmann constant and activation energy, respectively. The Arrhenius fit is shown by the solid red line in Fig. \hyperref[Fig. 2]{2}(a). From the fit, $\varepsilon_{act}$ is found to be $\approx$ 32 meV, suggests Ta$_2$Ni$_3$Te$_5$ is a narrow band-gap semiconductor, consistent with the theoretical  \cite{Guo2021} and experimental study on single crystal \cite{Yang2023}. In the mid-temperature region, the variable range hopping (VRH) is an appropriate model for explaining the temperature dependence of conductivity \cite{Ren2010, Shklovskii1984}. This model can be expressed as $\sigma \propto exp(T/T_0)^{-1/4}$,  where $\sigma$ is longitudinal conductivity, $T$ is temperature, and $T_0$ is the characteristic temperature of the material related to localization length ($\xi$) and density of states ($N(E_f)$). Fig. \hyperref[Fig. 2]{2}(c) presents a fit, shown in a red solid line, with the VRH model of the high-temperature conductivity data. A deviation from the fit suggests the presence of parallel conduction paths at low temperatures, consistent with observations for other topological materials \cite{Xu2014, Xia2012, Chen2010}. In the low-temperature region, there is a change of slope in the temperature dependence of resistivity, and a peak-shaped anomaly can be seen at $\approx$ 7 K, below which the resistivity shows metal-like temperature dependence. A recent study on Ta$_{2}$Pd$_{3}$Te$_{5}$ have revealed a similar temperature dependence for band gap and thus associated charge transport \cite{Wang2023}. In addition, a saturation like feature in resistivity was observed for Ta$_{2}$Pd$_{3}$Te$_{5}$ at low temperature regime \cite{Tian2022}

The origin of the metal-like resistivity has further been probed by temperature-dependent magnetotransport measurements. Fig. \hyperref[Fig. 2]{2}(d) shows temperature-dependent resistivity, normalized with respect to the resistivity at 300 K ($\rho(300)$), at various applied magnetic fields. A significant change in the nature of low-temperature transport behavior is observed under the influence of applied magnetic fields, shown in Fig. \hyperref[Fig. 2]{2}(d). The resistivity increases and the anomaly shifts towards higher temperatures with increase in the strength of the applied field. The relative increase in the low-temperature resistivity is less significant above the magnetic field of 4 T. Such temperature and field-dependent resistivity behavior strongly supports the presence of topological surface states (TSS). These TSSs are metallic in nature and are known to have conducting carriers of high mobility showing profound field dependent resistivity. In the temperature range of 8 K to 13 K, the magnetoresistance (for the applied field of 0.5 T) has decreased as compared to resistance in the zero applied field. Such behavior at low applied field is consistent with that observed in isostructural Ta$_2$Pd$_3$Te$_5$ \cite{Tian2022}. It is important to mention, that due to polycrystalline specimen, the anisotropic properties are averaged and the observed features of surface states can be associated with any particular crystal-axis. 

\begin{figure}[t!]
\includegraphics[width=1.0\columnwidth,angle=0,clip=true]{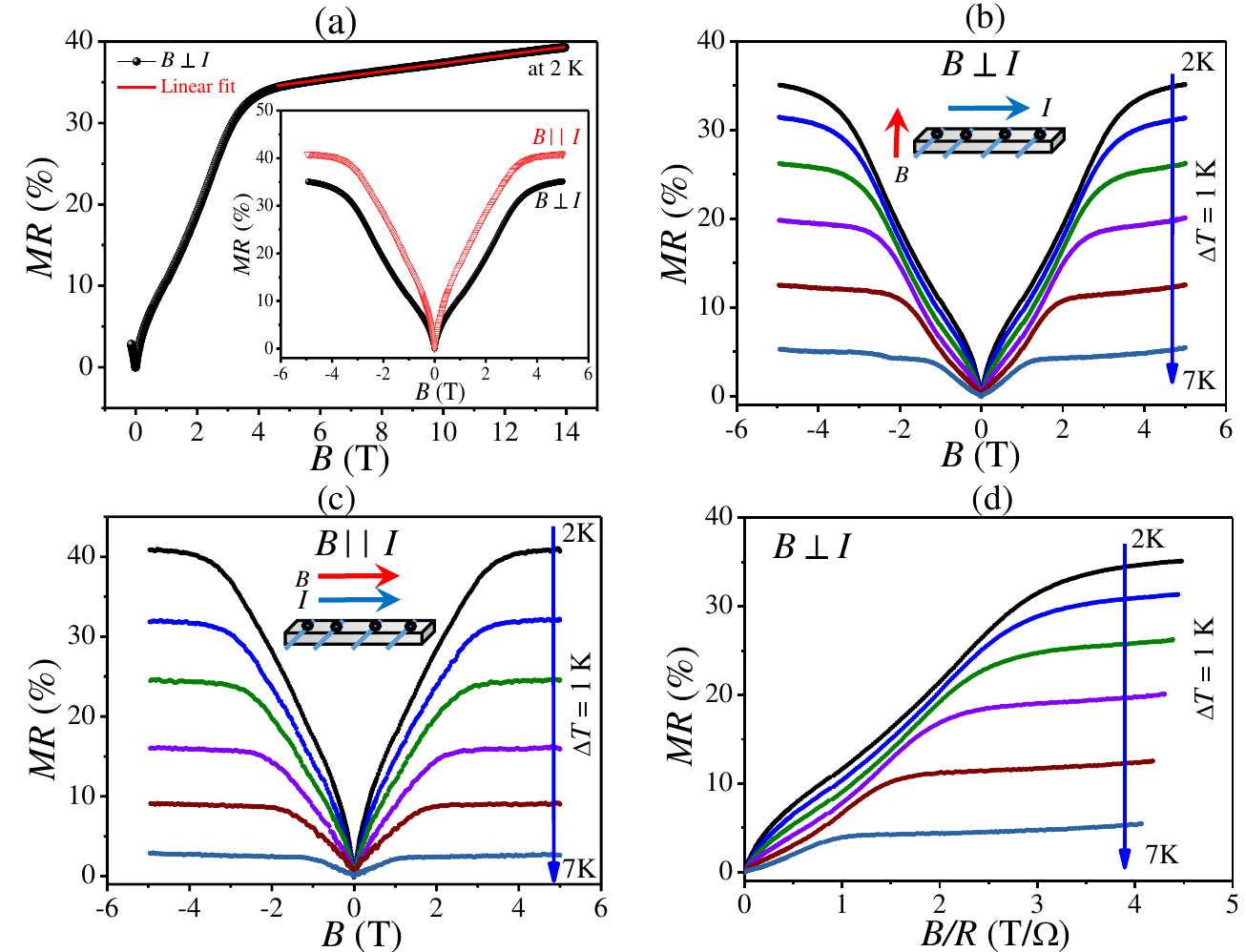}
\caption{(Color online) (a)MR$(\%)$ as a function of applied magnetic field up to 14 T at 2 K, a linear fit is shown at higher field range (from 5 T to 14 T) by the red line, inset shows MR$(\%)$ as a function of applied magnetic field ($\pm$ 5 T) perpendicular and parallel to the current direction at 2 K. (b) MR$(\%)$ at various temperatures from 2K to 7K with $\Delta T$ = 1 K, in range of applied field from -5T to 5T perpendicular to current direction. (c) MR$(\%)$ at various temperatures from 2K to 7K with $\Delta T$ = 1 K, in the applied field ($\pm$ 5 T) parallel to the current direction. (d) Kohler's plot corresponding to magnetic field perpendicular to the current direction.}
\label{Fig. 3}
\end{figure}

To further explore the nature of the electrical transport at low temperature, the magnetoresistance (MR) measurements have been carried out in the presence of an applied magnetic field parallel and perpendicular to the direction of the current. 
The schematic for both configurations has been provided in the supplementary.
The MR is defined by the relation MR$(\%)= \frac{[R(B)-R(0)]}{R(0)}\times{100}$. Where $R(B)$ and $R(0)$ are the resistances in the presence of finite and zero external magnetic fields, respectively. Fig. \hyperref[Fig. 3]{3}(a) presents MR$(\%)$ as a function of the applied magnetic field measured at 2 K. The isothermal MR measurements have been performed at various temperatures. Here, we have shown the MR data at temperatures below 7 K and MR at higher temperatures ($T>7 K$) are presented in the supplementary with analysis and description. A large ($\sim 40\%$) and non-saturating MR has been found for applied magnetic field up to 14 T. The non-saturating nature of the magnetoresistance has been demonstrated by the linear fit of high magnetic field data as shown in the red solid line. Although, apart from the surface states, the observation of linear MR is associated with other possible origins like (i) open Fermi surface (in case of metals) and (ii) impurity scattering (as observed in AgTe) and persists up to high temperature and applied field \cite{Abrikosov2000}. In our case, the unusual shape ($\Bar{}\text{V} \Bar{}$) of MR rules out its origin from impurity scattering (see Fig. \hyperref[Fig. 3]{3}(a) ). The large MR is evident in the limited low temperature ($T<$15 K) range, and diminished in high temperature and suggesting the surface states as more prominently contributing to linear MR. It is further supported by spin-momentum locking observed from isothermal magnetization data, discussed later. Classically, the longitudinal MR shows parabolic dependence with applied magnetic fields due to cyclotronic MR and saturates after a critical field limit \cite{Assaf2013, Hu2008}. Inset of Fig. \hyperref[Fig. 3]{3}(a) shows the MR measured in parallel and perpendicular configuration at 2 K up to an applied magnetic field of $\pm$5 T. Here perpendicular ($B \perp I$) and parallel ($B \parallel I$) configurations stand for the relative direction between applied magnetic field and current. There is a sharp cusp-like nature close to zero applied magnetic field, which has its origin from constructive quantum interference of time-reversed electronic paths, termed weak anti-localization (WAL). The mechanistic details of the WAL effect have been discussed in the appendix section. The Fig. \hyperref[Fig. 3]{3}(b) and (c) show the MR at several temperatures from 2 K to 7 K ($\Delta T = 1$ K), respectively in perpendicular and parallel configuration of the applied magnetic field, demonstrating the weak anti-localization and non-saturating nature. It has been established that the phenomenon of WAL is observed in the diffusive transport regime \cite{Grbi2008}. WAL is a quantum behavior of charge carriers and has a pronounced effect only at low temperature. Due to the decrease in phase coherence length at higher temperatures, the cusps expand and eventually disappear as the temperature rises. 
The non-saturating linear MR for high magnetic fields and the WAL effect for low magnetic fields can be attributed to the nontrivial topological nature of Ta$_2$Ni$_3$Te$_5$. 


The presence of electron-phonon ($e-p$) interaction is confirmed by Kohler's plot shown in Fig. \hyperref[Fig. 3]{3}(d), which is used for the relative study of MR measured at different temperatures. The cyclotron frequency and the relaxation time play key roles in the relative change in isothermal magnetoresistivity expressed in a relation given by $\frac{\Delta\rho(B)}{\rho(B)}= f\left(\frac{B}{\rho(0)}\right)$, where $\rho(B)$ is resistivity at an applied field of strength B, $\rho(0)$ is resistivity at zero field, and $\Delta\rho(B)$ = $\rho(B)- \rho(0)$. The MR$(\%)$ as a function of $B/R$ (applied magnetic field/resistance) is represented in Fig. \hyperref[Fig. 3]{3}(d). The MR curves at various temperature do not overlap with each other and thus violate Kohler's rule, suggesting significant electron-phonon interaction. 


In thin films and single crystals of topological systems, it has been observed that the conduction takes place via both the surface and bulk of the material even at considerably low temperature \cite{Zhang2019, Singh2017}. Systems having bulk as well as surface contribution in conductivity are studied with orientation-dependent (angle between applied magnetic field $(B)$ and current$(I)$) magnetotransport studies, to analyze the surface and bulk contributions, considering them as independent conduction paths \cite{Zhang2012, Zhang2011, Zhang2019}. Variation in magnetoresistance with orientation indicates the 2D nature of conduction, which is intertwined with the topology of electronic Fermi surfaces. In the case of materials with strong SOC, the weak anti-localization (WAL) effect is linked with bulk bands; it has been discussed with its origin from electron-electron interaction \cite{Lu2011, Wang2011}. Due to the polycrystalline nature of our sample, the WAL effect and nature of MR are comparable in both the parallel and perpendicular configurations (Fig. \hyperref[Fig. 3]{3}), which confirms the contribution from bulk conducting states along with surface states in MR. The sole purpose of applying the magnetic field in two different directions, (1)
parallel and (2) perpendicular to the current direction, is to investigate possible chiral anomaly as observed in weyl-semimetals. The chiral anomaly for the parallel applied magnetic field and current as negative MR and might be observed irrespective of crystal direction and thus can be probed within the polycrystalline specimen. From our positive MR data, we have confirmed the absence of chiral anomaly. A detailed analysis has further been done for the magnetoresistance data measured in the perpendicular configuration, where the applied magnetic field is perpendicular to the direction of current ($B \perp I$).

\section{perpendicular magnetotransport}

WAL effect along with linear MR are signatures of topologically non-trivial electronic band structures \cite{Zhang2012}. At low magnetic fields, WAL-like correction in magnetoconductivity can be well explained with Hikami-Larkin-Nagaoka  (HLN) equation \cite{Zhang2012, Hikami1980}. The HLN equation is mainly applicable in the diffusive regime and was derived considering the metallic nature of the specimen \cite{Grbi2008, Hikami1980}. The WAL effect is a quantum correction in magnetoconductance, having a phase coherence length ($l_{\phi}$), which is defined as the length traveled by a wave before it gets dephased due to scattering. The phase coherence length ($l_{\phi}$) is related to phase coherence time ($\tau_{\phi}$) by the relation, $l_\phi = 1/\sqrt{D\tau_\phi}$, where $D$ is the diffusion constant. 
In topological materials, the dominance of the contribution from the phase coherence over the contribution from the spin-orbit coupling is significant at a low applied magnetic field and decreases with an increase in the applied magnetic field and temperature; consequently, the WAL effect is more pronounced at low applied magnetic field \cite{Zhang2012}. A characteristic field $B_\phi$, analogous to the magnetic field, is defined such that it corresponds to phase coherence length $l_\phi$, having a relation as $B_\phi= h/8\pi el_\phi^2$, where $h$ is Plank's constant and $e$ is the electronic charge, \cite{Lu2011}. The HLN equation relates the relative change in magnetoconductance, $\Delta G (B)$ = $G(B) - G(0)$, with applied magnetic field ($B$) given by:

\begin{equation}
\label{hln1}
\Delta G(B) = \frac{\alpha e^2}{2\pi h} \left[\psi\left(\frac{1}{2} + \frac{B_\phi}{B}\right) - \rm {ln} \left(\frac{B_\phi}{B}\right)\right],
\end{equation}

where $G(B)$ and $G(0)$ are conductance in the presence and absence of applied magnetic field, respectively; $\psi$ is a digamma function, $\alpha$ is the prefactor which corresponds to the number of conduction channels. In an ideal 2D WAL effect, the value of $\alpha$ is equal to -1, the deviation from which corresponds to surface-bulk coupling and/or a larger number of conduction paths. 

\begin{figure}[t!]
\includegraphics[width= 1\columnwidth,angle=0,clip=true]{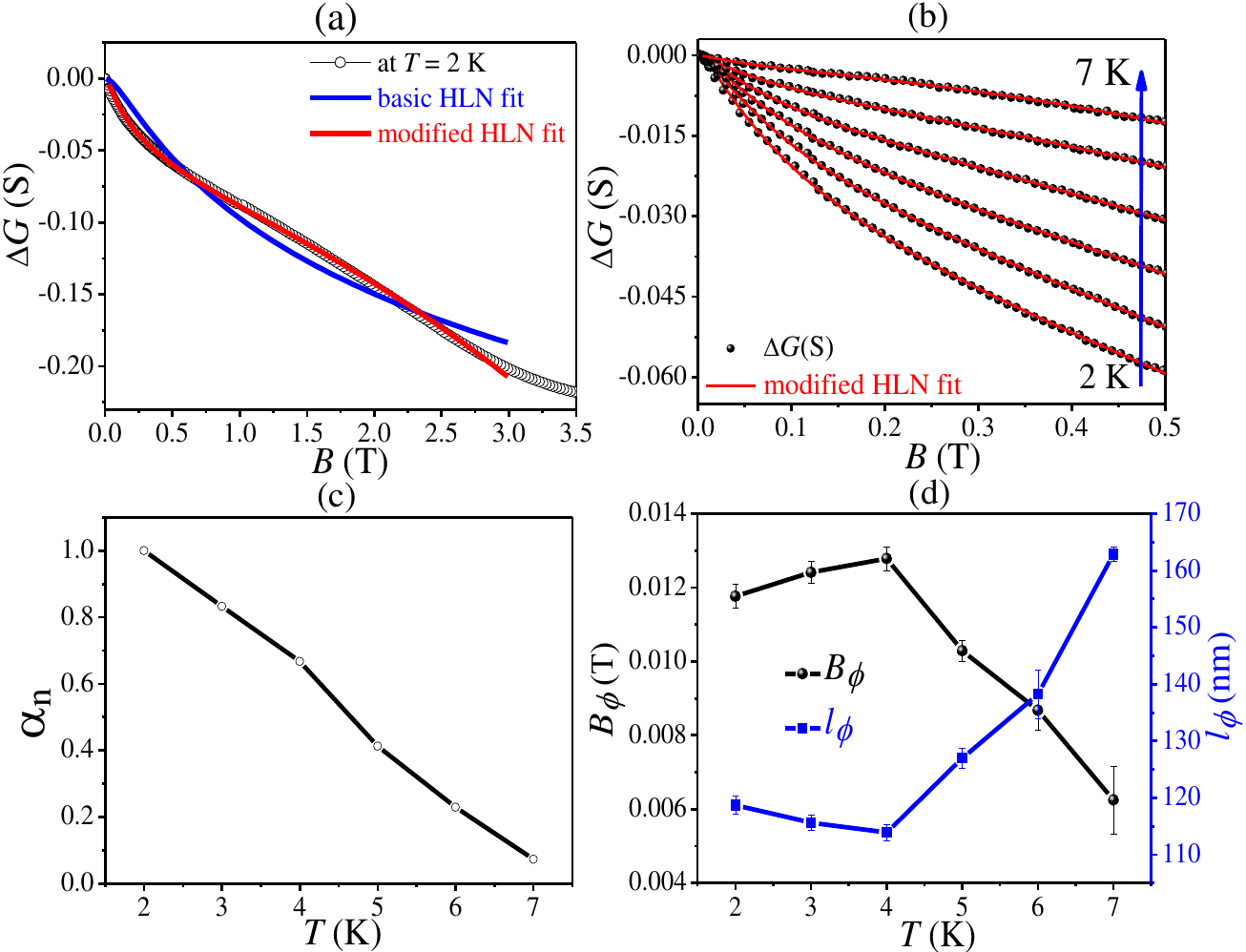}
\caption{(Color online) (a) Magnetoconductance as a function of the applied field, fitting curves for simple and Modified-HLN equations. Modified-HLN equation is providing a better fit of data (b) Magnetoconductance fitted with modified-HLN equation at different temperatures from 2 K to 7 K with $\Delta T$ = 1 K. (c) Variation of normalized $\alpha$ with temperature. (d) Variation of $B_\phi$ and $l_\phi$ with temperature.}
\label{Fig. 4}
\end{figure}

The contribution from both bulk and surface states in the WAL effect can be present up to the lowest measured temperature (2 K) \cite{Nomura2007, Dey2016, Zhang2012, Lu2011}. Therefore, considering both surface and bulk contributions in the modified HLN equation is essential for the appropriate treatment of the MR data \cite{Dey2016}. In the modified-HLN equation (Eq. \hyperref[hln2]{2}), all other classical and quantum corrections to magnetotransport except the phase coherence are approximated to have $B^2$ dependence \cite{Assaf2013}, and can be expressed as:

\begin{equation}
\label{hln2}
\Delta G (B) = \frac{\alpha e^2}{2\pi h} \left[\psi\left(\frac{1}{2} + \frac{B_\phi}{B}\right) - \rm {ln} \left(\frac{B_\phi}{B}\right)\right] - cB^2,
\end{equation}

where $c$ is the coefficient of the $B^2$ dependance. Close to zero magnetic field and at low temperatures, the contribution from spin-orbit coupling is significantly less than that from phase coherence; hence the analysis of the low-temperature magnetoconductivity data has been carried out for applied magnetic field up to 0.5 T. In Fig. \hyperref[Fig. 4]{4}(a) we have shown the fit of the magnetoconductivity data with the basic HLN (shown in blue solid line) and modified HLN (shown in red solid line) equations. As evident from Fig. \hyperref[Fig. 4]{4}(a), the modified equation, which includes spin-orbit scattering, inelastic scattering, and classical dependence, provides a better fit for a larger range of applied magnetic fields while fitting with the basic HLN equation shows considerable deviation from the data. This implies the low-temperature magnetoconductivity data, and hence the WAL effect, have contributions from the surface as well as bulk conducting states.

Magnetoconductance $vs.$ applied magnetic field data at various temperatures, shown in Fig. \hyperref[Fig. 4]{4}(b), have been fitted with the modified-HLN equation to obtain the temperature dependence of the parameters $\alpha$ and $B_\phi$. In our case, the value of $\alpha$ is significantly  larger ($\alpha$ at 2 K = 3895), $\sim 10^3$ times, than its value for the 2D limit ($\alpha$ = 1 for conduction through each single surface state). This is due to the availability of a larger number of conduction paths in 3D polycrystalline bulk specimens. In literature, the existence of such large values of $\alpha$ have been reported for topological insulators Bi$_2$Se$_3$ \cite{Checkelsky2009}, Bi$_2$Te$_3$ \cite{Shrestha2017}, also for single crystals of Bi-based half-Heusler alloys like YPtBi \cite{Pavlosiuk2016}, LuPdBi \cite{Xu2014}, ScPtBi \cite{Hou2015}, and for Cu doped CaAuAs which is a topological semimetal \cite{Malick2022}. The temperature dependence of $\alpha_n$ (= $\alpha(T)/\alpha(2K)$) is shown in Fig. \hyperref[Fig. 4]{4}(c), where $\alpha_n$ is the normalized value of $\alpha$ at various temperatures with respect to the value of $\alpha$ at 2 K. The monotonic decrease in the value of $\alpha$ with an increase in temperature suggests a relative decrease in the conduction through the surface states. 
From the fit, phase coherence length, $l_\phi$ has been estimated to have values in the range of 110-165 nm, within the measured temperature interval of 2-7 K. These values are similar to the reported values for a single crystal of Ta$_2$Pd$_3$Te$_5$ \cite{Tian2022}. According to Nyquist, dephasing due to electron-electron scattering phenomenon, should have dependence given by relation $l_\phi \propto T^{-1/2}$, for ideal 2D systems \cite{Breznay2012, Wu2012}. In our case, $l_\phi$ shows an irregular variation with temperature, it decreases initially and then increases with an increase in temperature, as shown in Fig. \hyperref[Fig. 4]{4}(d), thus not showing ideal temperature dependence expected for a 2D system and suggested to have the contribution of electron-electron interactions along with electron-phonon interactions in dephasing mechanism. Variation of $l_\phi$ with temperature indicates a considerable contribution from bulk charge carriers in the conductivity of the 3D specimen, which is accompanied with multiple scatterings involved in electron dephasing at elevated temperatures, supported by Kohler's plot study (Fig. \hyperref[Fig. 3]{3}(d)). 
In order to have a complete understanding of these interactions and scattering mechanisms in Ta$_2$Ni$_3$Te$_5$, experimental studies on thin films and single crystals are desirable.


\section{Magnetic properties}

\begin{figure}[t!]
\includegraphics[width=.98 \columnwidth,angle=0,clip=true]{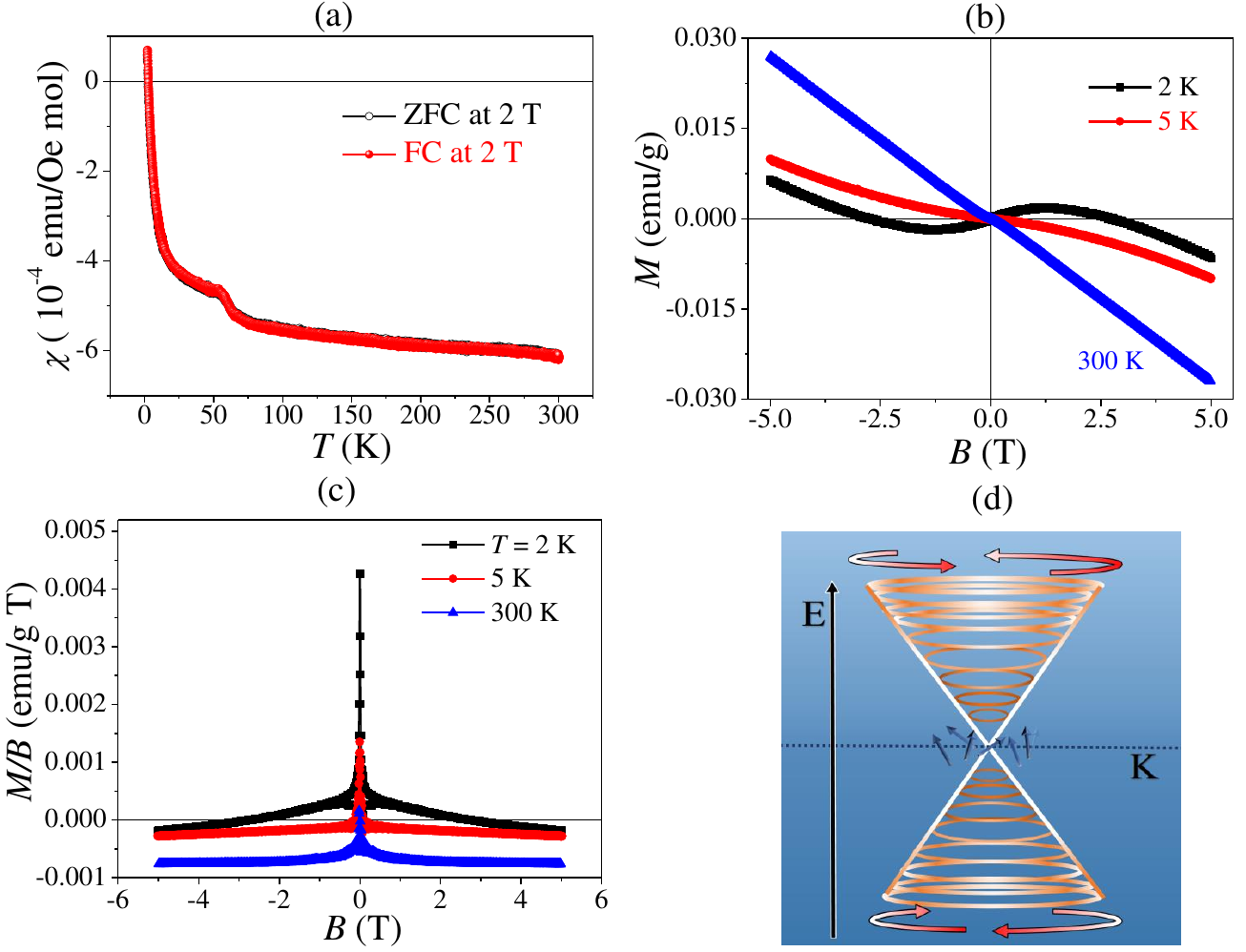}
\caption{(Color online) (a) Magnetic susceptibility of Ta$_2$Ni$_3$Te$_5$ in Zero field cooled and field cooled protocol as a function of temperature for the applied field of 2 T. (b) Isothermal magnetization at 2, 5 and 300 K. (c) Magnetic susceptibility ($M/B$) as a function of the applied magnetic field at different temperatures, cusp at the low applied field is clearly visible. (d) The schematic E-K diagram of Dirac-like electronic state, enriched in topological materials}
\label{Fig. 5}
\end{figure}

Temperature-dependent DC magnetic susceptibility measurements have been performed in zero field cooled (ZFC) and field cooled (FC) protocols with the applied magnetic field of 20 kOe, shown in Fig. \hyperref[Fig. 5]{5(a)}, in the temperature range of 2-300 K. 
Both the ZFC and FC susceptibility data show diamagnetic nature with an increase in susceptibility at low temperature below 10 K, and eventually become positive below $\approx$ 2.5 K, suggesting existence of low-temperature paramagnetic component along with the diamagnetic signal. There is a small hump close to 55 K, which possibly arises due to short-range interactions. 
Recent study on Ta$_2$Ni$_3$Te$_5$ shows overall paramagnetic behavior of the magnetic susceptibility measured at applied magnetic field of 1 kOe \cite{Yang2023}. To further investigate the magnetic nature of Ta$_2$Ni$_3$Te$_5$, field-dependent istothermal magnetization study has been performed.
%


Isothermal magnetization measurements at different temperatures have been carried out in the applied magnetic field ranging from -5 T to +5 T, as shown in Fig. \hyperref[Fig. 5]{5}(b). At \sout{5} \textcolor{red}{2} K, initially the magnetization increases with increase in the applied magnetic field as a paramagnetic response, then decreases beyond a certain magnetic field and becomes negative at higher applied fields. The isothermal magnetization at 5 K and 300 K show similar behavior with relatively larger diamagnetic response. Fig. \hyperref[Fig. 5]{5}(c). Significant paramagnetic response in the low-field regime is evident from the data, which gradually decreases with increasing temperature. This paramagnetism is associated with the nontrivial band topology and termed as Berry paramagnetism which originates from the zeroth Landau level of the nontrivial electronic band structure in topological materials \cite{Ji2021, Moll2016}. The high-field negative magnetization can be rationalized with diamagnetism generated by orbiting motion of a large number of trivial electrons of valence band, which overcomes the positive response of paramagnetic behavior at higher magnetic fields \cite{Moll2016}. 
The paramagnetic nature at the low-field regime is evident in Fig. \hyperref[Fig. 5]{5}(c), which shows $M/B$ $vs.$ $B$. We found cusp-like low-field paramagnetic susceptibility that grows over high-field diamagnetic platform. Such a signature in the magnetic susceptibility originates from spin-momentum locking, which leads to the alignment of the electronic spins, close to the Dirac point, along the applied magnetic field. This behavior of magnetization with the applied magnetic field is reminiscent of the Dirac-like dispersion of fermions, schematically depicted in the inset of Fig. \hyperref[Fig. 5]{5}(d), and is a signature of topological surface states \cite{Singha2017, Zhao2014}. The existence of Dirac-like surface states in Ta$_2$Ni$_3$Te$_5$ is supported by theoretical evidences for nontrivial band topology of this material \cite{Guo2022, Yang2023}. 

Presence of Dirac-like dispersion of fermions provides an explanation for the increase in the $\chi(T)$ data at low temperature (Fig. \hyperref[Fig. 5]{5}(a)). The low-temperature paramagnetic component, therefore, can be attributed to Berry-paramagnetism. Furthermore, the temperature-dependent magnetic susceptibility measured at 1 kOe by Yang $et$ $al.$ \cite{Yang2023} is quite consistent with our data measured at 20 kOe, since it is evident from the isothermal magnetization study that the low-field paramagnetic response changes to diamagnetic signal for higher magnetic fields. 



Both the polycrystalline and single crystal phases of ZrTe$_5$, a topological material, show similar cusp-like behavior in the low field region of the $M/B$ $vs.$ $B$ plot \cite{Hooda2018, Pariari2017}. Similar cusp-like isothermal susceptibility is also evident for topological materials Bi$_2$Se$_3$ and Bi$_2$Te$_3$, and found to be independent of bulk carrier concentrations \cite{Zhao2014}. In case of Bi$_2$Te$_3$, the occurrence of WAL \cite{Chiu2013} and spin-momentum locking \cite{Zhao2014} in the same low temperature range suggests that both are intertwined phenomena. In our case of Ta$_2$Ni$_3$Te$_5$, the existence of WAL effect in the magnetotransport data and the spin-momentum locking phenomenon in the isothermal magnetization are comparable in terms of the range of the temperature and the applied magnetic field. This suggests that both of these phenomena are not mutually exclusive, but rather correlated with each other. Detailed study is required to further investigate the mutual dependence of the WAL effect and spin-momentum locking in topologically non-trivial system Ta$_2$Ni$_3$Te$_5$.

\begin{figure}[!t]
\includegraphics[width=.98 \columnwidth,angle=0,clip=true]{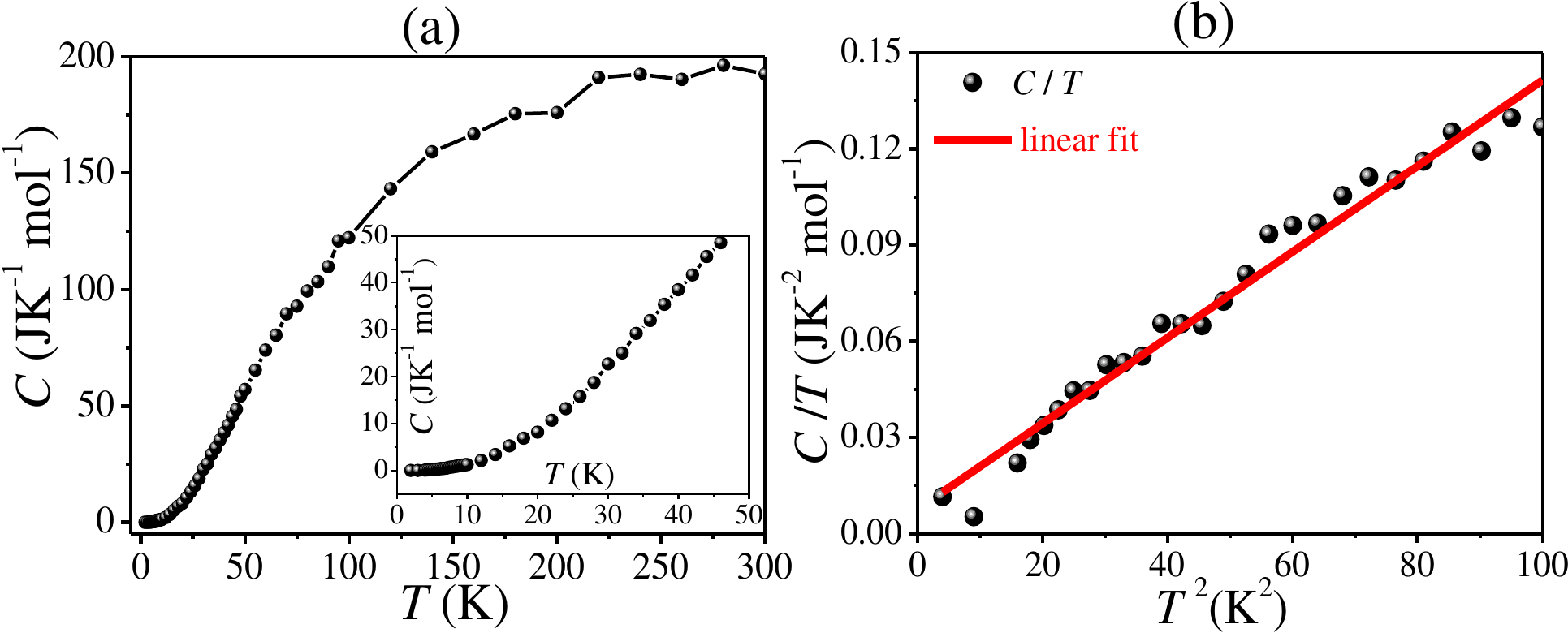}
\caption{(Color online) (a) Specific heat of Ta$_2$Ni$_3$Te$_5$  as a function of temperature at zero magnetic field, inset shows expanded view of $C$ $vs.$ $T$ data in low-temperature regime (b) $C/T$ $vs.$ $T^2$ plot in low-temperature regime ($T \leq$ 10 K); fit with $C/T= \gamma + \beta T^2$ is shown by the red solid line.}
\label{Fig. 6}
\end{figure}

\section{Specific heat}

Specific heat study of Ta$_2$Ni$_3$Te$_5$ was carried out over a temperature range of 2-300 K at zero magnetic field as shown in Fig. \hyperref[Fig. 6]{6}(a), and inset shows the expanded view of the data at low-temperature scale. 
There is no significant anomaly present in the specific heat signal, 
suggests absence of any structural transition or long-range magnetic ordering, further indicating that the semiconducting to metal-like transition in the resistivity data at low temperature can be attributed to the non-trivial topological nature of the system.
In the range of temperature 2-10 K, specific heat data can be analyzed for having electronic and phonon contributions, related by the equation

\begin{equation}\label{CT}
C(T)/T = \gamma+\beta T^2,
\end{equation}

where $C$ is the specific heat capacity, $T$ is temperature, $\gamma$ is Sommerfeld coefficient, and $\beta$ is phonon specific-heat coefficient. From the fit of the temperature-dependent specific heat data with Eq. \hyperlink{Theta}{3}, as shown in Fig. \hyperref[Fig. 6]{6}(b), the Sommerfeld coefficient and the phonon specific-heat coefficient are estimated to be $\gamma$ = 7.55 $\pm$ 2.46 mJ K$^{-2}$ mol$^{-1}$ and $\beta$ = 1.34 $\pm$ 0.05 mJ K$^{-4}$ mol$^{-1}$, respectively.
Phonon specific heat coefficient, $\beta$, is related to the Debye temperature ($\theta_D$) of the material by the relation

\begin{equation}\label{Theta}
\theta_D = \left[\frac{12}{5\beta}\pi^4nN_Ak_B\right]^{1/3},
\end{equation}

where $n$ is the number of atoms in the formula unit, $N_A$ is the Avogadro number, and $k_B$ is the Boltzmann constant. Debye temperature is estimated to be 244 $\pm$ 3 K, by using Eq. \hyperlink{Theta}{4}. Sommerfeld coefficient, $\gamma$ is related to the density of states $N(E_F)$ of the system close to the Fermi level, which is calculated by using the relation 

\begin{equation}\label{Gamma}
\gamma = [\pi^2 k_B^2 N(E_F)]/3.
\end{equation}

$N(E_F)$ is found to be 3.2 $\pm$ 1.0 states/eV f.u., which is close to the theoretically calculated value of $N(E_F)$ ($\approx$ 3) for isostructural Ta$_2$Pd$_3$Te$_5$ \cite{Guo2021}. 

\section{Conclusions} 

In summary, we have synthesized polycrystalline Ta$_2$Ni$_3$Te$_5$ and studied its magnetotransport, magnetic, and specific heat properties. At high temperature, Ta$_2$Ni$_3$Te$_5$ is found to be a narrow gapped semiconductor with $\varepsilon_{act}\approx$ 32 meV. There is a semiconductor-to-metal transition at $\approx$ 7 K, indicating parallel conduction through bulk and surface conducting states. From the field-dependent study, large MR($\%$) has been obtained accompanied with  WAL-like correction in magnetotransport at low magnetic field, which is further analyzed with the modified HLN equation. The inclusion of the $B^2$-dependent term in the HLN equation to achieve a better fit of magnetoconductivity data indicates a significant contribution from bulk states along with surface states. We obtained large values for the parameter $\alpha$, indicating that conduction occurs through multiple conducting channels down to 2 K, as expected in a polycrystalline sample. The irregular variation of $l_\phi$ with temperature indicates dephasing occurring through multiple scattering mechanisms. This is further supported by the non-overlapping feature in Kohler's plot. Bulk contribution in the origin of the WAL effect in magnetoresistivity is further confirmed by observing the similar extent of MR in parallel (B$\parallel$I) and perpendicular (B$\perp$I) configurations. The presence of spin-momentum locking is supported by the paramagnetic susceptibility cusp (Berry paramagnetism) at low applied field prominent at low temperature in isothermal magnetization measurements, which suggests the presence of topological surface states with Dirac-like dispersion. Further, the specific heat data confirm the absence of any structural phase transition or long-range magnetic ordering, suggesting the low-temperature characteristics of the system are attributed to its nontrivial band topology. The presence of WAL-like correction along with large MR$(\%$) in magnetotransport studies, and paramagnetic cusp in isothermal magnetic susceptibility provide signatures of nontrivial electronic band topology in Ta$_2$Ni$_3$Te$_5$. To gain extensive understanding of its non-trivial topological properties, more detailed experimental studies on Ta$_2$Ni$_3$Te$_5$ single crystals are required. This work presents a comprehensive study of the physical properties of polycrystalline Ta$_2$Ni$_3$Te$_5$, and offers a promising platform to explore the nontrivial topology in ternary layered chalcogenides for quantum spin Hall insulator-based applications.


\section*{appendix}

In the Dirac-cone states of nontrivial topological systems, the spin-momentum locking generated by strong spin-orbit coupling (SOC) leads to a Berry phase that reduces backscattering and produces quantum interference.
If the phase coherence length is substantially higher than the mean free path, the electrons preserve phase coherence even after elastically scattering several times, resulting in weak localization and weak antilocalization in the time-reversed interference path. When a modest magnetic field is applied, the interference disappears, and WL and WAL produce a cusp-like negative and positive magnetoresistance, respectively. There is a greater likelihood of scattering events occurring between two time-reversed routes in lower-dimensional systems, therefore, the WL and WAL effects are more prominent in thin film studies and layered materials.

\section*{ACKNOWLEDGMENTS}

The authors acknowledge the Central Research Facility (CRF), IIT Delhi for experimental facilities. We thank Dr. Dinesh Dixit for his help and support during the resistivity, magnetotransport, and specific heat measurements. PKM acknowledges the Council of Scientific $\&$ Industrial Research (CSIR), India for fellowship.

\bibliographystyle{apsrev4-2}
\bibliography{reference}

\end{document}